\documentstyle[sprocl]{article}

\input{psfig}
\def \x {{\bf x}}
\def \y {{\bf y}}
\def \A {{\bf A}}
\def \B {{\bf B}}
\def \D {{\bf D}}

\def \r {{\bf r}}

\def\bbox {}

\def\be{\begin{equation}}
\def\ee{\end{equation}}
\def\bea{\begin{eqnarray}}
\def\eea{\end{eqnarray}}
\begin{document}

\title{Hybrids and Quark Confinement
\footnote{Presented at the Conference on
Perspectives in Hadronic Physics, Trieste, Italy and QCD97, Montpellier, France.}}

\author{E.S. Swanson}

\address{Dept of Physics, North Carolina State University,
Raleigh\\ NC 27695-8202, USA\\ and \\
Jefferson Laboratory, 12000 Jefferson Ave, Newport News, VA, 23606, USA}

\maketitle\abstracts{
It has become traditional to assume that the Dirac structure of the 
phenomenological quark confinement potential is scalar $\otimes$ scalar. 
We use the heavy quark expansion of the Coulomb gauge QCD Hamiltonian and the
Flux Tube model to demonstrate that this is true but in the {\em effective} 
sense only. The demonstration
contains some surprises: confinement is actually vector $\otimes$ vector and
it is nonperturbative mixing between ordinary
and hybrid $Q \bar Q$ states which generates the scalar-like spin dependent
potential at order $1/m_q^2$.  Thus the existence of hybrids is crucial to
establishing well-known spin splitting phenomenology. 
Finally, the resolution also indicates that the gluonic degrees of freedom
in a hybrid must be of a collective nature.
}

\section{Introduction} 

Although it has been postulated for more than 30 years, 
the phenomenon of quark confinement remains an enigmatic feature of QCD.
Quenched lattice gauge theory and heavy quark phenomenology indicate that the
static ($m_q >> \Lambda_{QCD}$) long range potential should be linear with a 
slope 
of $b \approx 0.18\, {\rm GeV}^2$. The order $1/m_q^2$ quark-antiquark 
long range  spin-dependent (SD) structure is also of interest and is the
subject of this paper.
In a model approach, ${Q\bar Q}$ interactions are typically derived from 
a nonrelativistic reduction of a relativistic current-current interaction

\begin{equation}
 V_{conf} = \int j_\Gamma(\x) K(\x -\y) j_\Gamma(\y).
\end{equation}

\noindent
In the heavy quark, nonrelativistic limit, this reduces to 

\begin{equation}
V_{conf} \rightarrow \epsilon(r) + V_{SD} + \ldots
\end{equation}

\noindent
where the ellipsis denotes the spin independent interaction at ${\cal O}(1/m^2)$
and terms of higher order in $1/m$. The spin dependent interaction may be 
parameterized as\cite{EF} 

\begin{eqnarray}
V_{SD}(r) &=& \left( {\bbox{\sigma}_q \cdot {\bf L}_q \over 4 m_q^2} -
{\bbox{\sigma}_{\bar q} \cdot {\bf L}_{\bar q} \over 4 m_{\bar q}^2} \right) \left( {1\over r}
{d \epsilon \over d r} + {2 \over r} {d V_1 \over d r} \right) + 
\left( {\bbox{\sigma}_{\bar q} \cdot {\bf L}_q \over 2 m_q m_{\bar q}} - 
{\bbox{\sigma}_q \cdot {\bf L}_{\bar q} \over 2 m_q m_{\bar q}} \right) 
\left( {1 \over r} {d V_2 \over d r} \right) \nonumber \\
&& 
+ {1 \over 12 m_q m_{\bar q}}\Big( 3 \bbox{\sigma}_q \cdot \hat {\bf r} \, 
\bbox{\sigma}_{\bar q}\cdot \hat {\bf r} -   \bbox{\sigma}_q\cdot  
\bbox{\sigma}_{\bar q} \Big) V_3(r) 
+ {1 \over 12 m_q m_{\bar q}} \bbox{\sigma}_q \cdot \bbox{\sigma}_{\bar q} 
V_4(r). \label{VSD}
\end{eqnarray}
Here $\epsilon=\epsilon(r)$ is the static potential,  
$r= |{\bf r}_q - {\bf r}_{\bar q}|$ is the ${\bar Q Q}$ separation 
and the $V_i=V_i(r)$ are potentials which depend on the nonperturbative
structure of QCD.
Assuming that the $V_i$ simply follow from Eq. (1) yields the results shown 
in Table I \cite{Grev}. Tensor and axial vector currents have spin-dependent
static potentials and are therefore not useful for phenomenology in this 
context. Even the derivation of these simple results is beset with ambiguities
and controversy. The reader is referred to the excellent review by Gromes in 
Ref. \cite{Grev}. Finally we note that covariance under Lorentz transformations
leads to a constraint between the SD potentials: 
$\epsilon(r) = V_2(r) - V_1(r)$, known as Gromes' relation\cite{G}.

\begin{table}[h]
\caption{Nonrelativisitic limits of Eq. (1). \label{tab:1}}
\vspace{0.4cm}
\begin{center}
\begin{tabular}{|c|c|c|c|c|c|}
\hline
 $\Gamma$ &  $\epsilon_\Gamma$ &  $V_1$ & $V_2$ & $V_3$ & $V_4$ \\
\hline
scalar &  $S$ &  $-S$ & 0 & 0 & 0 \\
vector &  $V$ &  0  & $V$ & $V'/r - V''$ & $2 \nabla^2 V$ \\
pseudoscalar &  0 &  0  & 0  & $P'' - P'/r$ & $\nabla^2 P$ \\
\hline
\end{tabular}
\end{center}
\end{table}

Twenty-two years ago Schnitzer\cite{schnitzer} examined splittings in 
P-wave heavy quarkonia in
an attempt to distinguish the possible spin dependence of the confinement
potential. Specifically, he assumed a vector Coulomb short range potential and
scalar or vector long range potentials. The ratio of P-wave splittings is then
given by

\begin{equation}
\rho_{\buildrel V \over S} \equiv {M(^3P_2) - M(^3P_1)\over M(^3P_1) - M(^3P_0)} = {\langle 
{16 \alpha_s\over 5 r^3} \pm A {b\over r}\rangle \over \langle {4\alpha_s 
\over r^3} \pm B {b \over 2 r}\rangle},
\end{equation}

\noindent
where $A = 14$ (1) and $B = 4$ (1) for the vector (scalar) case.
Experimentally, the splittings for the $J/\psi$ and $\Upsilon$ systems are
0.48 and 0.66 respectively. These are shown as horizontal lines in Fig. 1. 
The upper and lower curves are $\rho_V$ and $\rho_S$ respectively (these have
not been convoluted with the wavefunctions). Since the $b\bar b$ wavefunction is
smaller than the $c \bar c$, the curve must drop with increasing $r$ to 
fit the data. Clearly only scalar confinement meets this criterion.
This conclusion has been supported by many other methods, including lattice
gauge theory\cite{michael}, Wilson loop calculations\cite{mbp,nora}, and flux 
tube models\cite{B}.

\begin{figure} 
\hbox{ \hfil \qquad \qquad \psfig{figure=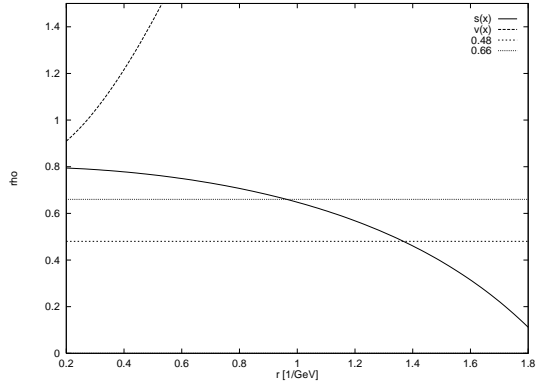,height=2in} \hfil}
\caption{Scalar (s) and vector (v) spin splittings vs. the 
quark-antiquark radius. \label{fig:1}}
\end{figure}

A consistent
picture of a QCD-generated effective scalar confinement interaction appears
to be emerging. It is therefore disconcerting that attempts to build  
models which are ``closer'' to QCD (typically relativistic models) seem to 
require vector confinement. For example,  
scalar Salpeter equations
do not have stable solutions\cite{PP} and it appears to be impossible to construct
a BCS-like vacuum of QCD when scalar confinement is assumed\cite{ssjc,ss}.
This is problematical if one wishes to dynamically generate constituent 
quark masses.
Furthermore, attempts at modeling chiral pions will be hindered by the
explicit lack of chiral symmetry in a scalar interaction. Finally, baryons
anticonfine and colour singlet Tamm-Dancoff or RPA
bound states are infrared divergent in a scalar potential\cite{ss4}.

This issue was resolved in Ref. \cite{ss2} 
by first performing a Foldy-Wouthuysen 
reduction of the full Coulomb gauge Hamiltonian of QCD. This immediately
establishes that the Dirac structure of confinement for heavy quarks is of 
a timelike-vector nature.  
Furthermore, the spin dependent structure of the long range force is determined
by nonperturbative mixing of the $q\bar q$ states with hybrids. Finally, if
one assumes a flux tube-like configuration of the gluonic component of the 
hybrids,
a ``scalar'' spin-dependent potential emerges in a natural way.

\section{Heavy Quark Expansion of $H_{QCD}$}

Our starting point is the Coulomb gauge QCD Hamiltonian

\begin{equation}
H_{QCD} = \int d\x \psi^\dagger(\x)\left[-i\bbox{\alpha}\cdot \nabla + 
\beta m\right]\psi(\x) + H_{YM} + V_C 
- g \int d\x \psi^\dagger(\x) \bbox{\alpha}\cdot {\bf A}(\x) \psi(\x) 
\label{QCD}
\end{equation}

\noindent
where

\begin{equation}
V_C = {1\over 2} g^2\int d\x d{\bf y} \, {\cal J}^{-1} \rho^a(\x) V^{ab}({\bf x}, 
{\bf y}; A) {\cal J} \rho^b({\bf y}),\label{vc}
\end{equation}

\begin{equation}
V^{ab}(\x,\y;A) = \langle \x,a |(\nabla\cdot {\cal D})^{-1} 
(-\nabla^2) (\nabla\cdot {\cal D})^{-1} | \y,b \rangle, \label{vab}
\end{equation}

\noindent
and

\begin{equation}
H_{YM} =  {1\over 2}\int d\x \left[ {\cal J}^{-1}\bbox{\Pi}(\x) {\cal J} 
\bbox{\Pi}(\x) + {\bf B}^2(\x) \right].
\end{equation}

\noindent
The degrees of freedom are the transverse gluon field $\A=\A^a{\rm T}^a$,
its conjugate momentum  $\bbox{\Pi}=\bbox{\Pi}^a {\rm T}^a$, and the quark field in the 
Coulomb gauge. The Faddeev-Popov determinant is written as
${\cal J} = \mbox{Det}[\nabla\cdot {\cal D}]$, with ${\cal D}^{ab} = \nabla\delta^{ab} 
- g f^{abc}\A^c$ being the  covariant derivative in the adjoint representation,
and the magnetic field is given by 
${\bf B}^a = \nabla\times \A^a + g f^{abc}\A^b \times \A^c$. The static 
interaction $V_C$ is the nonabelian analog of the Coulomb potential. It 
involves the full QCD 
color charge density which has both quark and gluon components,
\begin{equation}
\rho^a(\x) = \psi^\dagger(\x) {\rm T}^a \psi(\x) + f^{abc} \A^b(\x) \cdot
\bbox{\Pi}^c(\x).
\label{rho}
\end{equation} 

The most salient feature of the Coulomb gauge Hamiltonian is that all of
the degrees of freedom are physical. This makes it especially useful for
identifying the physical mechanisms which drive the spin splittings in
heavy quarkonia.

\subsection{The Foldy-Wouthuysen Transformation}

We proceed by performing a Foldy-Wouthuysen transformation on the QCD
Hamiltonian. This is done in complete analogy to the quantum mechanical
case where an operator is constructed which removes the interactions
between upper and lower components of the quark wave function order by order
in the inverse quark mass, 
except that the unitary transformation is now constructed in
Fock space.
 The resulting Hamiltonian  is given by 
 
\begin{eqnarray}
H_{QCD} \to H_{FW} &=& \int d {\bf x} \left( m_q h^\dagger({\bf x}) h({\bf x}) - 
m_{\bar q} \chi^\dagger({\bf x}) \chi({\bf x})\right) + H_{YM} + \nonumber \\
       && V_C + H_1 + H_2 + \ldots,  
\end{eqnarray}
\begin{eqnarray}
H_1 &=& {1 \over 2m_q} \int d{\bf x} h^\dagger({\bf x}) \left(\D^2 - g \bbox{\sigma} \cdot 
{\bf B} \right) h({\bf x}) - ( h \rightarrow \chi; m_q \rightarrow m_{\bar q}), \\
H_2 &=& {1 \over 8 m_q^2} \int d{\bf x} h^\dagger({\bf x}) 
 g \bbox{\sigma}\cdot [{\bf E}, \times {\bf D}] 
h({\bf x}) + ( h \rightarrow \chi; m_q \rightarrow m_{\bar q}). 
\end{eqnarray}

\noindent
In this expression  $h = (1 + \beta) \psi/2$ and $\chi = (1 - \beta) \psi/2$ 
denote the upper and lower components of the quark wave function and correspond
to the annihilation and 
creation operators of the heavy quark and antiquark respectively. 
The ellipsis denotes terms which are either of $O(1/m^3)$ or are 
spin-independent at order 
$1/m^2$. Finally  $\D = i \nabla + g \A$ is the covariant derivative in the 
fundamental representation.

\subsection{The Static Potential}

To leading order in the quark mass the Hamiltonian describes two static, 
noninteracting quarks. 
At ${\cal O}(m^0)$ the Hamiltonian reduces to $H_0 = H_{YM} + V_C$.
The eigenstates of $H_0$ may be labeled by
 the quark and antiquark coordinates and by an index which classifies the adiabatic state
of the gluonic degrees of freedom, $n_r$,
\begin{equation}
H_0 |n_r; \r_q \r_{\bar q} \rangle = \epsilon_n(r) |n_r; \r_q 
\r_{\bar q}\rangle.
\end{equation}
Note that we have made explicit the
dependence of the gluonic degrees of freedom on the position of the quarks, $r$.

The corresponding eigenenergies, $\epsilon_n(r)$  may be identified with the 
Wilson loop potentials calculated on 
the lattice. Thus for example, $\epsilon_0(r)$ is the Coulomb plus linear 
potential seen long ago. Static hybrid states are collectively denoted
$\vert n_r;\r_q \r_{\bar q}\rangle$ with  $n_r  \ne 0$. In  recent 
studies \cite{pm}  (see Fig. 2)
 the  lowest lying 
adiabatic hybrid potential, $\epsilon_1(r)$ has been evaluated.

\begin{figure} 
\hbox{ \hfil \qquad \qquad \psfig{figure=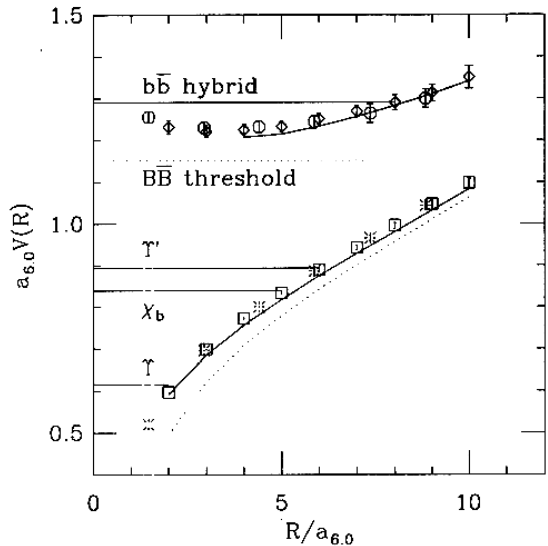,height=3.in} \hfil}
\caption{{$\epsilon_0(R)$ (squares) and $\epsilon_1(R)$ (circles and diamonds). 
From Ref. 14. \label{fig:2}}}
\end{figure}

While both $H_{YM}$ and $V_C$ may contribute to the linearly rising potential energy seen 
on the lattice, it is clear that the quarks may
only interact with the flux tube via the nonabelian Coulomb interaction. Therefore the Dirac 
structure of confinement corresponds to $\gamma_0\otimes\gamma_0$  from the product of color 
 charge densities (see Eqs.~(\ref{vc}) and ~(\ref{rho})). 
As stressed in the
Introduction, this appears to be at odds with 20 years of quark model 
phenomenology. Since the phenomenology is based on spin splittings, it 
will be instructive to examine the $1/m^2$ perturbative 
corrections to the static potential.

\subsection{Spin-dependent Potentials}

The spin-dependent first order correction to the static potential is given by
$\delta\epsilon_n^{(1)}(r) = \langle n_r; \r_q \r_{\bar q}\vert H_2 
\vert n_r;\r_q \r_{\bar q}\rangle$.
The order $1/m$ term is not considered because $\D^2$ is not spin-dependent and
the matrix element of $\bbox{\sigma}\cdot\B$ vanishes.
Evaluating this 
yields the standard classical plus Thomas precession spin orbit interaction

\begin{equation}
\delta \epsilon_n^{(1)} = 
\left( {\bbox{\sigma}_q \cdot {\bf L}_q \over 4 m_q^2} -
{\bbox{\sigma}_{\bar q} \cdot {\bf L}_{\bar q} \over 4 m_{\bar q}^2} \right)
 {1\over r}
{d \epsilon_n \over d r}.
\end{equation}

\noindent
Thus the first term in Eq.~(\ref{VSD}), generalized to any adiabatic potential and therefore true for both 
ordinary and hybrid $\bar Q Q$ states, 
is reproduced. 
Further SD corrections occur at second order in perturbation theory in
$H_1$. This is shown diagramatically in Fig. 3. As indicated in the 
figure, $H_1$ contains 
two operators, $D^2$ and $\sigma\cdot \B$ which may act on quark or antiquark 
lines. The shaded blobs represent mixing via intermediate states which must
be heavy quark hybrids. In Fig. 3a $\sigma\cdot\B$ acts on a single quark line
and therefore does not yield a spin-dependent contribution to $V_{SD}$.
Fig. 3b, alternatively, yields $V_3$ and $V_4$; while Figs. 3c and 3d 
give $V_1$ and $V_2$ respectively.

Thus, the application of the Foldy-Wouthuysen transformation to the 
Coulomb gauge QCD 
Hamiltonian shows that the spin-dependent effective potential may be simply 
interpreted as nonperturbative mixing with hybrid states. 
This makes it clear that it is possible for
nonperturbative dynamical physics to generate an effective spin-dependent
interaction which mimics a scalar interaction. 
Actually demonstrating this requires that the matrix elements be evaluated.

\begin{figure}
\psfig{figure=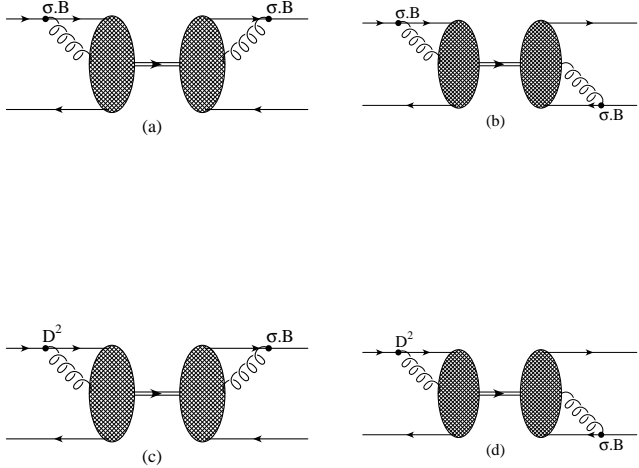,height=2.5in}
\caption{Contributions to the second order energy shift. 
\label{fig:3}}
\end{figure}

\section{Model Evaluation of the Spin-dependent Potentials}

Before proceeding to a model evaluation of the matrix elements we note that 
it is possible to make some general statements on their expected
properties. The results of lattice gauge theory make it clear that a 
flux tube-like configuration of glue exists between static quarks. If one
thinks of this as a localized object with an infinite number of degrees of
freedom, then it is apparent that $V_2$ must evaluate to zero. This is because
the electric field operator creates a local excitation in
the flux tube at position $\r_q$ (see Fig. 3d). This must then be 
de-excited at $\r_{\bar q}$ by the magnetic field operator. 
However, the two operators become decorrelated because infinitely many 
degrees of freedom intervene. 
Similarly, the long range portions of  $V_3$ and
$V_4$ both vanish. Thus, by Gromes' relation,  the only nonzero long
range interaction must be given by $V_1 = - \epsilon$. This is precisely 
the situation required for ``scalar" confinement. It is therefore entirely 
plausible
that an effective scalar confinement is generated by nonperturbative mixing 
with hybrids. Furthermore,  the structure of the spin-dependent terms depends 
crucially on the nature of the ground state gluonic degrees of 
freedom and clearly favors a collective rather than a  single particle picture
of them.

These simple expectations are borne out in an explicit model 
calculation\cite{ss2} where the Flux Tube Model of Isgur and Paton \cite{IP} 
was used to evaluate the relevant matrix elements. The interested reader
is referred to Ref. \cite{ss2} for details.  
Evaluating the spin dependent potentials in the Flux Tube Model 
requires explicit
expressions for the electric and magnetic fields in a flux tube. It was 
therefore
necessary to extend the Flux Tube Model somewhat. Here I take the opportunity
to correct some typos in Ref. \cite{ss2}.

The commutation relation between the electric and magnetic fields is

\begin{equation}
[E^a_i(\x), B^b_j(\y)] = i \epsilon^{ijk} \nabla^k_y \delta(\y-\x) \delta^{ab} +
{\cal O}(g), \label{com}
\end{equation}

\noindent
which implies that the electric and magnetic field operators may be defined 
in terms of the transverse string coordinate $\chi$ as

\begin{equation}
E^a_\lambda(n) \sim {\kappa \over a^3}( \chi^a_\lambda(n+1) - 
\chi^a_\lambda(n)) \label{ef}
\end{equation}

\noindent
and 

\begin{equation}
B^a_\lambda(n) = {-i \over \kappa a} {\partial \over \partial 
\chi^a_{\bar \lambda}(n)}. \label{bf}
\end{equation}

\noindent
Here $\kappa = a \sqrt{b_0}$ and the $\epsilon^{ijk}$ of the commutator in 
Eq.~(\ref{com}) is taken into account by the index 
$\bar \lambda$,  $\chi_{\bar x}(n) =  \chi_y(n)$ and 
$\chi_{\bar y}(n)= - \chi_x(n)$. 
 
\noindent
This corrects several sign errors in Ref. \cite{ss2}. Note also that Eq. 38 of
that paper should be in terms of $\bar \lambda$ not $\lambda$. None of the
results are changed.

Finally, evaluating the matrix elements from the perturbative expansion yields 

\begin{equation}
V_1(r) = -{g^2 \over 2 a^2} C_F r
\end{equation}

\noindent
and 

\begin{equation}
V_2(r) = \lim_{N\rightarrow \infty} {g^2 \over 2 a^2} C_F {r \over N}.
\end{equation}

\noindent
In the strong coupling limit one has $b = g^2/(2 a^2) C_F$ so that the 
anticipated expression for $V_1$ emerges in a natural way.  
Furthermore, $V_2$ approaches zero like $1/N$; this is
also true for $V_3$ and $V_4$. 
The latter point is illustrated in Fig. 4 where the 
correlation of electric and magnetic fields versus separation along the 
flux tube is shown. 
As expected the fields become completely decorrelated as
the number of intervening degrees of freedom becomes large. Notice that this 
implies
that a constituent gluon model of hybrids would not have been able to produce
an effective scalar interaction.

\begin{figure}
\psfig{figure=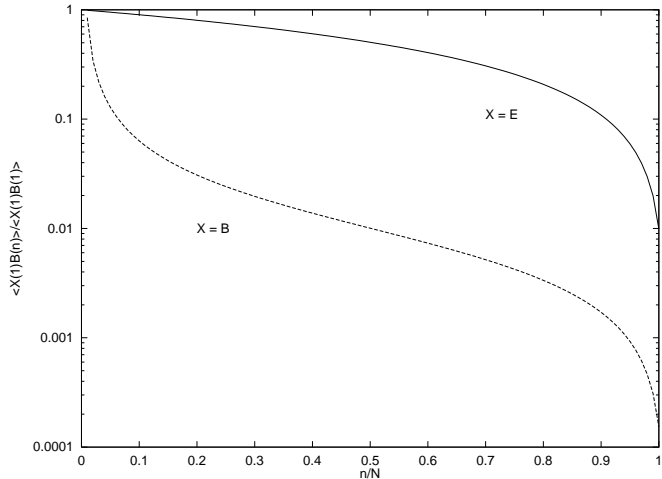,height=2.5in}
\caption{Field Correlation Functions (for N=100). 
\label{fig:4}}
\end{figure}

\section{Conclusions}

Spin splittings and lattice calculations indicate that confinement is scalar in
nature. This conflicts with many relativistic models of QCD which require vector
confinement. For example, a chirally symmetric interaction is needed if 
pseudo-Goldstone pions  and spontaneous chiral symmetry breaking  
 are to be generated dynamically.  
 Furthermore it appears to be impossible
to build a stable vacuum with a scalar kernel. We have examined
this issue with the heavy quark limit of the Coulomb gauge QCD Hamiltonian. 
This approach is physically intuitive and is simpler to interpret and implement than
methods based on the Wilson loop. We found that the static confinement potential
must indeed be a Dirac timelike vector. Effective scalar interactions are generated at
order $1/m^2$ by nonperturbative mixing with hybrid states. 

We have argued that the long range spin-spin ($V_3$ and $V_4$) and the vector-like
spin-orbit potentials ($V_2$) should all be zero since they involve field 
correlation functions evaluated between quark and antiquark. This statement
follows by assuming that the gluonic degrees of freedom collapse into a 
flux tube-like configuration, as shown by lattice gauge theory. Alternatively, the scalar-like
spin-orbit potential ($V_1$) is proportional to the matrix element of the electric
and magnetic fields evaluated at the same point and hence is expected to be 
nonzero. Explicit calculations of the relevant matrix elements were carried out in 
the Flux Tube Model. The model was extended to include color degrees of
freedom and to map the chromoelectric and chromomagnetic fields to flux tube 
phonon operators. The results obtained were in agreement with our general 
arguments and with Gromes' relation. 

A consistent picture of
the Dirac structure of confinement has emerged. The static central potential is
timelike vector while the spin-dependent structure mimics the nonrelativistic
reduction of an effective scalar interaction. This implies that it is incorrect
to employ a scalar confinement kernel when doing calculations with light quarks.
Note however that it would be acceptable to use scalar confinement when working 
explicitly in the chiral symmetry broken phase, {\it i.e.}, 
with constituent quarks in the nonrelativistic limit. The work presented here also
implies that a constituent gluon picture of hybrids will yield incorrect results
for certain observables. For example, $V_1$ and $V_2$ would be of comparable
magnitude in a constituent gluon model. In general, these types of models
must fail when 
nonlocal properties of the gluonic configuration are considered. Alternatively, 
it is possible that they
perform quite well when evaluating global properties of gluonics such as 
the hybrid spectrum.

\section*{Acknowledgements}
ES acknowledges the financial support of the DOE under grant 
DE-FG02-96ER40944.

\end{document}